\documentclass[prb,onecolumn,superscriptaddress,showpacs]{revtex4}
\usepackage{graphicx,stmaryrd,amsmath,amssymb,feynmf}

\begin{document}

\title{Impurity scattering in unconventional density waves: non-crossing approximation for arbitrary scattering rate}

\author{Andr\'as V\'anyolos}
\email{vanyolos@kapica.phy.bme.hu}
\affiliation{Department of Physics, Budapest University of Technology and Economics, 1521 Budapest, Hungary}
\author{Bal\'azs D\'ora}
\affiliation{Department of Physics, Budapest University of Technology and Economics, 1521 Budapest, Hungary}
\author{Kazumi Maki}
\affiliation{Department of Physics and Astronomy, University of Southern California, Los Angeles CA 90089-0484, USA}
\author{Attila Virosztek}
\affiliation{Department of Physics, Budapest University of Technology and Economics, 1521 Budapest, Hungary}
\affiliation{Research Institute for Solid State Physics and Optics, PO Box 49, 1525 Budapest, Hungary.}
\date{\today}

\begin{abstract}
We present a detailed theoretical study on the thermodynamic properties of impure quasi-one dimensional
unconventional charge-, and spin-density waves in the framework of mean-field theory. The impurities are of the
ordinary non-magnetic type. Making use of the full self-energy that takes into account all ladder-, and
rainbow-type diagrams, we are able to calculate the relevant low temperature quantities for arbitrary scattering
rates. These are the density of states, specific heat and the shift in the chemical potential. Our results
therefore cover the whole parameter space: they include both the self-consistent Born and the resonant unitary
limits, and most importantly give exact results in between.
\end{abstract}

\maketitle

\section{Introduction}
In the past few decades many papers have been published dealing with the mean-field theoretical study of impurity
scattering in either density waves (DW) or superconductors (SC). It has been pointed out earlier by several
authors,\cite{lra,schuster,patton} that non-magnetic impurities should have a pair-breaking effect on conventional
density waves, similar to the one caused by paramagnetic impurities in SCs, found first by Abrikosov and Gor'kov
(AG).\cite{abrikosov} In this connection, the original AG theory,\cite{abrikosov-book,parks-book} developed
specifically for the treatment of dilute, randomly distributed magnetic impurities in $s$-wave BCS superconductors,
was later successfully applied with minor modifications for example in excitonic
insulators,\cite{zittartz1,zittartz2} and in the context of conventional charge and spin density waves with
constant gap.\cite{roshen,viro-sdw-optics} All these studies however, particularly because of the assumed dilute
concentration, considered the impurity effects on the level of second order self-consistent Born approximation, and
retained only the first nontrivial self energy insertion in the proper self energy caused by impurities. On the
other hand, in case of strong scatterers, higher-order diagrams in the self energy might be of importance,
especially at low temperatures in view of the anomalous behavior of the normal state (Kondo
effect).\cite{maki-anomalous,griffin}

After the discovery of unconventional superconductors, the extension of the field of density waves to DWs with
wave-vector dependent gap $\Delta(\mathbf{k})$ (termed unconventional) arose in a natural way. Unconventional
density waves (UDW), either spin or charge, have the interesting property, that though a robust thermodynamic phase
transition is clearly seen in experiments, they lack any spatial modulation of charge or spin density due to the
zero average of the gap over the Fermi surface. Consequently, the phase transition cannot be caught in the act with
conventional means like x-ray or NMR. This state of affairs is usually referred to as the ``hidden order'' in recent
literature.\cite{laughlin} One of the main reasons of interest on UDWs arises from high-$T_c$ superconductors,
where one of the competing models in the pseudogap phase is the $d$-density wave state, a kind of two-dimensional
unconventional charge density wave (UCDW).\cite{congjun1,kim,valenzuela} The relevance of UDWs in general is
further emphasized by recent studies such as: the mysterious low temperature micromagnetic phase of some heavy
fermion compounds seem to be described rather well by unconventional spin density wave (USDW),\cite{ikeda,attila}
the pseudogap phase of (TaSe$_4$)$_2$I is explained, at least on qualitative level, by a coexisting CDW+UCDW
order,\cite{nemeth,tase4} and finally the ground state of the transition metal dichalcogenid 2H-TaSe$_2$ was
attributed to an $f$-wave UCDW lately.\cite{castro-neto} As to the quasi-one dimension, which is the natural place
of occurence of DWs, one of the most likely candidate for possessing some kind of CDW of unconventional type in its
low temperature phase is the organic conductor $\alpha$-(BEDT-TTF)$_2$KHg(SCN)$_4$. In the past few years this salt
has been investigated extensively,\cite{andres,mori,kartsovnik,basletic,fujita,pouget} and recently the
experimental findings of these latter works have been explained rather well by a quasi-one dimensional UCDW,
including the threshold electric field,\cite{balazs-threshold,balazs-magneticthreshold,balazs-imperfect} the
angular dependent magnetoresistance,\cite{balazs-ucdw-et,balazs-eurlett} and the magnetothermopower and Nernst
effect\cite{balazs-ucdw-mtpnernst}.

Impurity scattering in UDWs and unconventional SCs has been investigated many times from different aspects, but the
theoretical formulations were limited to include either the weak scattering Born or the resonant unitary
limit.\cite{balazs-born,balazs-unitary,sigrist} The main outcome of these studies is that the presence of
non-magnetic impurities leads to the destruction of ground state. The present work is partly motivated by the
surprising discovery that the Ni impurities enhance the energy gap of $d$-wave DW in underdoped YBCO.\cite{pimenov}
In light of this, our aim in this paper is to explore the intermediate regime as well by making use of the full
self energy in the non-crossing approximation. We give a detailed analysis of the thermodynamic properties of UDWs,
without making any constraint on the strength of the forward and backscattering parameters. As we shall see, our
general formulas suggest that non-magnetic impurities always suppress density wave ground state. The enhancement of
the pseudogap in YBCO by magnetic Ni impurities is proposed to originate from a different mechanism discussed in a
separate publication.\cite{balazs-nickel} Calculations similar to this has been carried out for heavy fermion
superconductors\cite{hirschfeld} and $d$-density waves\cite{d-density-wave-bdg,d-density-wave-impurity}. To the
best of our knowledge however, this question in quasi-one dimensional UDWs has not been addressed so far.

The article is organized as follows: in Section 2 we present the formalism necessary to treat impurity scattering
in the non-crossing approximation in unconventional density waves. In Section 3 we calculate the relevant
thermodynamic quantities in the low temperature phase. Section 4 is dedicated to the analysis of the quasiparticle
density of states, and finally Section 5 is devoted to our conclusions.

\section{Formalism}
We begin with the mean-field Hamiltonian describing pure quasi-one dimensional density
waves\cite{gruner-book,balazs-sdw}
\begin{equation}
H_0=\sum_{\mathbf{k},\sigma}\xi(\mathbf{k})(a_{\mathbf{k},\sigma}^+a_{\mathbf{k},\sigma}-a_{\mathbf{k-Q},\sigma}^+a_{\mathbf{k-Q},\sigma})+\Delta_\sigma(\mathbf{k})a_{\mathbf{k},\sigma}^+a_{\mathbf{k-Q},\sigma}+\Delta_\sigma^*(\mathbf{k})a_{\mathbf{k-Q},\sigma}^+a_{\mathbf{k},\sigma},\label{meanfieldham}
\end{equation}
where $a_{\mathbf{k},\sigma}^+$ and $a_{\mathbf{k},\sigma}$ are the creation and annihilation operators for an
electron in a single band with momentum $\mathbf{k}$ and spin $\sigma$. The best nesting vector is given by
$\mathbf{Q}=(2k_F,\pi/b,\pi/c)$, and the $\mathbf{k}$ sum in Eq.~\eqref{meanfieldham} is restricted to the reduced
Brillouin zone as usual: $k_x$ runs from 0 to $2k_F$, with $k_F$ being the Fermi wave vector. Our system is based on
an orthorhombic lattice with lattice constants $a,b,c$ towards directions $x,y,z$. The system is highly anisotropic,
the kinetic energy spectrum of the particles linearized around the Fermi surface reads as
\begin{equation}
\xi(\mathbf{k})=v_F(k_x-k_F)-2t_b\cos(bk_y)-2t_c\cos(ck_z),
\end{equation}
and the quasi-one dimensional direction is the $x$ axis. For (U)CDWs the density wave order parameter
$\Delta_\sigma(\mathbf{k})$ is even, while for (U)SDW systems it is an odd function of the spin index. In the case
of conventional density waves the order parameter is constant on the Fermi surface.\cite{gruner-book} As opposed to
this, for unconventional condensates it depends on the perpendicular momentum and has different values at different
points on the Fermi surface. The precise $\mathbf{k}$-dependence is determined by the matrix element of the
electron-electron interaction through the gap equation.\cite{balazs-sdw}

The interaction of the electrons with non-magnetic impurities is described by the
Hamiltonian\cite{gorkov,balazs-born,balazs-unitary}
\begin{align}
H_1&=\frac1V\sum_{\mathbf{k,q},\sigma, j}e^{-i\mathbf{qR}_j}\Psi_\sigma^+(\mathbf{k+q})U(\mathbf{R}_j)
\Psi_\sigma(\mathbf{k}),\label{imphamilton}\\
U(\mathbf{R}_j)&=
\begin{pmatrix}
U(0) & U(\mathbf{Q})e^{-i\mathbf{QR}_j} \\
U^*(\mathbf{Q})e^{i\mathbf{QR}_j} & U(0)
\end{pmatrix},\label{impurity}
\end{align}
where $V$ is the sample volume, $\mathbf{R}_j$ is the position of the $j$th impurity atom on the lattice, and the
momentum space is divided into $\mathbf{k}$ and $\mathbf{k-Q}$ subspaces (right- and left-going electrons) by
introducing the spinor
\begin{equation}
\Psi_\sigma(\mathbf{k})=
\begin{pmatrix}
 a_{\mathbf{k},\sigma}\\
 a_{\mathbf{k-Q},\sigma}
\end{pmatrix}.\label{spinor}
\end{equation}
We have neglected the small $\mathbf{q}$-dependence of the matrix elements (i.e. the Fourier components of the
impurity potential) in Eq.~\eqref{impurity}. This is because we are dealing with quasi-one dimensional density wave
systems whose Fermi surface consists of two almost parallel sheets at the points $\pm k_F$. Consequently, two
relevant scattering amplitudes can be distinguished: the forward $U(0)$ and the backward scattering parameter
$U(\mathbf{Q})$, respectively. We believe that we can capture the essence of physics with this approximation. This
is indeed the case, at least as long as only thermodynamics is concerned. Investigating the effect of pinning, and
the sliding density wave however, requires the inclusion of the small $\mathbf{q}$
contribution.\cite{balazs-threshold,balazs-magneticthreshold}

Now, with all this, the electron thermal Green's function for the interacting system using Nambu's notation is
defined as\cite{parks-book}
\begin{equation}
G_\sigma(\mathbf{k},i\omega_n)=-\int_0^\beta d\tau\langle T_\tau\Psi_\sigma(\mathbf{k},\tau)
\Psi_\sigma^+(\mathbf{k},0)\rangle_H e^{i\omega_n\tau},
\end{equation}
where the Green's function is chosen to be diagonal in spin indices, and the subscript $H$ refers to the total
Hamiltonian $H=H_0+H_1$ with which the expectation value should be taken. The propagator for the pure system
without impurities is readily evaluated as
\begin{equation}
G_\sigma^0(\mathbf{k},i\omega_n)=-\frac{i\omega_n+\xi(\mathbf{k})\rho_3+\rho_1\text{Re}\Delta_\sigma(\mathbf{k})-\rho_2\text{Im}\Delta_\sigma(\mathbf{k})}{\omega_n^2+\xi(\mathbf{k})^2+|\Delta_\sigma(\mathbf{k})|^2},\label{puregreen}
\end{equation}
where $\rho_i$ $(i=1,2,3)$ stand for the Pauli matrices acting on the space of left-, and right-going electrons.

From now on we will drop the spin indices since they are irrelevant for our discussion on thermodynamic properties,
and our results apply to both unconventional spin-, and charge-density waves as well. Furthermore, we choose the
momentum dependence of the order parameter appearing in the zeroth order Green's function as
$\Delta(\mathbf{k})=\Delta e^{i\phi}\sin(bk_y)$, where the phase $\phi$ is unrestricted due to
incommensurability. Though the impurities we shall shortly introduce in the system are able to pin down the
phase,\cite{balazs-threshold,balazs-magneticthreshold} performing the usual impurity average restores translational
invariance and consequently the phase degree of freedom. Thus, for brevity we fix the phase to be $\phi=0$ and
without the loss of generality consider a real order parameter. As to the specific $\mathbf{k}$-dependence, had we
chosen a gap with cosine function or with the $z$ component of the wave vector would not alter our results.

\begin{figure}
\begin{fmffile}{graph1}
\begin{equation}
 \parbox{10mm}{
  \begin{fmfgraph}(20,40)
  \fmfsurroundn{e}{4}
  \fmf{phantom}{e1,v}
  \fmf{dashes}{e2,v}
  \fmfv{decor.shape=cross,decor.size=4thick}{e2}
  \fmf{phantom}{v,e3}
  \fmf{phantom}{v,e4}
  \end{fmfgraph}}
\quad + \quad
 \parbox{30mm}{
  \begin{fmfgraph}(60,40)
  \fmfsurroundn{e}{4}
  \fmf{fermion}{e1,e3}
  \fmf{dashes}{e2,e1}
  \fmf{dashes}{e2,e3}
  \fmfv{decor.shape=cross,decor.size=4thick}{e2}
  \fmf{phantom}{e1,e4}
  \fmf{phantom}{e3,e4}
  \end{fmfgraph}}
\quad + \quad
 \parbox{30mm}{
  \begin{fmfgraph}(60,40)
  \fmfsurroundn{e}{4}
  \fmf{fermion}{e1,v}
  \fmf{fermion}{v,e3}
  \fmffreeze
  \fmf{dashes}{e2,e1}
  \fmf{dashes}{e2,v}
  \fmf{dashes}{e2,e3} 
  \fmfv{decor.shape=cross,decor.size=4thick}{e2}
  \fmf{phantom}{e1,e4}
  \fmf{phantom}{v,e4}
  \fmf{phantom}{e3,e4}
  \end{fmfgraph}}
\quad + \quad \dots
\nonumber
\end{equation}
\end{fmffile}
\vspace*{-10mm}
\caption{Non-crossing self energy corrections due to impurity scattering. The solid line denotes the electron
propagator $G$, while the dashed line is for the electron-impurity interaction. Dashed lines coming from the same
cross represent successive scattering of the electron on the same impurity.\label{diagrams}}
\end{figure}
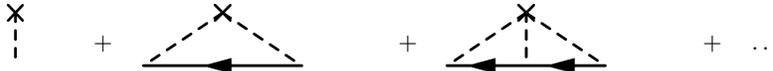

Let us now turn our attention to the self-energy $\Sigma(\mathbf{k},i\omega_n)$ caused by the non-magnetic
impurities. In diagrammatic language, the non-crossing approximation (NCA) retains only those diagrams where no two
interaction lines cross each other. A simple subset of such diagrams are usually referred to as the ladder-type
diagrams describing successive scattering processes on the same and single impurity atom. In case the particle
lines appearing in these diagrams are not the bare $G^0$ but the dressed propagators $G$, then all the rainbow-type
diagrams are included as well, and the NCA becomes full and self-consistent, see Fig.~\ref{diagrams}. The coupled
equations valid for UDW are\cite{balazs-born}
\unitlength 0.5mm

\begin{align}
G^{-1}(\mathbf{k},i\omega_n)&=G_0^{-1}(\mathbf{k},i\omega_n)-\Sigma,\label{dyson}\\
\Sigma&=n_i\frac{U(0)-g(U(0)^2-|U(\mathbf{Q})|^2)}{1-2gU(0)+g^2(U(0)^2-|U(\mathbf{Q})|^2)},\label{selfenergy}
\end{align}
where $g=(2V)^{-1}\sum_\mathbf{k}\text{Tr}G(\mathbf{k},i\omega_n)$ is the local Green's function, $n_i$ is the
impurity concentration, and according to the ignorance of small $\mathbf{q}$-dependence of scattering amplitudes
(see Eq.~\eqref{impurity} and subsequent argument), the self energy turns out to be momentum independent
$\Sigma(\mathbf{k},i\omega_n)\equiv\Sigma(i\omega_n)$.
We shall note here that the impurity concentration should be small enough in order for the NCA to be applicable and
reliable. Quantitatively it means that the self energy should be much smaller than the bandwith $W=2v_Fk_F$,
otherwise the contribution of crossing diagrams cannot be neglected and retaining the non-crossing class of
diagrams only becomes inadequate. Moreover, at large enough concentration the standard independent average over the
impurity positions on the lattice breaks down (for the impurity average performed on $\Sigma$, see
Ref.~\onlinecite{balazs-born}). We will see shortly however, that because of pair-breaking, we have a natural limit
on $n_i$ which is of the order of 1\%. Staying in this region of concentrations should be within the bounds.

The ansatz-solution $G$ of the closed pair of equations above reads as
\begin{equation}
G(\mathbf{k},i\omega_n)=-\frac{i\tilde\omega_n+\xi(\mathbf{k})\rho_3+\Delta(\mathbf{k})\rho_1}{\tilde\omega_n^2+\xi(\mathbf{k})^2+\Delta(\mathbf{k})^2},\label{exactgreen}
\end{equation}
where, in contrast to conventional systems with constant gap, only the Matsubara frequency gets
renormalized.\cite{balazs-born,balazs-unitary} The order parameter remains unaffected due to the unconventional
nature of the gap with vanishing Fermi surface average. This is also known for quasi-one dimensional
superconductors.\cite{suzumura} With the aid of the explicit expression in Eq.~\eqref{exactgreen}, $g$ is calculated
easily, and the renormalization condition for the Matsubara frequency follows as
\begin{align}
i\tilde\omega_n&=i\omega_n+\delta\mu-\Sigma,\label{alapegyenlet}\\
g&=-N_0\frac{i\tilde\omega_n}{\sqrt{\tilde\omega_n^2+\Delta^2}}K\left(\frac{\Delta}{\sqrt{\tilde\omega_n^2+\Delta^2}}\right),\label{kisg}
\end{align}
where $N_0=1/(\pi v_Fbc)$ is the normal state density of states per spin, $K(z)$ is the complete elliptic integral
of the first kind, and we have introduced a shift in the chemical potential $\delta\mu$ by hand in
Eq.~\eqref{alapegyenlet}. This we need in order to keep the particle number conserved in the system, since the
presence of impurities on a macroscopic scale causes the particles to either flow in or out of the system depending
on the sign and magnitude of the forward and backward scattering parameters.

Impurity scattering from the theoretical point of view has been investigated extensively in the literature. These
related works can be usually classified into one of two groups depending on the formalism used to capture the
strength of the scattering process. However, both approaches are limiting cases of the formalism introduced
here. In the second order Born approximation,\cite{roshen,maki-viro,balazs-born,balazs-unitary} applicable for weak
impurity scattering, the self energy simplifies to $\Sigma=n_iU(0)-n_ig(U(0)^2+|U(\mathbf{Q})|^2)$, and
$\delta\mu=n_iU(0)$. On the other hand, for strong scattering the unitary limit is widely
used,\cite{balazs-unitary,sun-maki,puch-maki,maki-puch-pwave} where $\Sigma=-n_i/g$ and $\delta\mu=0$.

The aim of the present work is to make use of the full self energy defined in Eq.~\eqref{selfenergy}. Without
making any constraints on the strengths of the scattering parameters $U(0)$ and $U(\mathbf{Q})$, we give a unified
theory of the full non-crossing approximation. With this general approach we are able to cover the whole parameter
space, to reobtain the well known formulas in the weak and strong scattering limits, and most importantly to
provide exact results in the intermediate regime. As we have already mentioned, similar calculations based on a
self-consistent $T$-matrix approximation have been performed for heavy fermion superconductors\cite{hirschfeld},
and then later for $d$-density waves\cite{d-density-wave-impurity} as well. In UDWs however this question is still
open.

\section{Thermodynamics of impure UDW}
\begin{figure}
\begin{center}
\includegraphics[width=8.6cm,height=6.5cm]{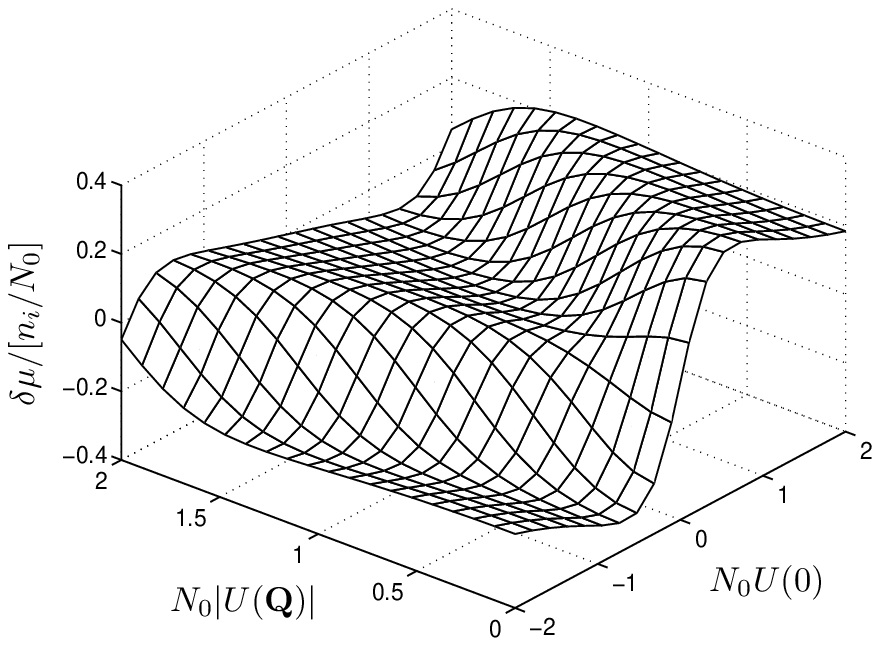}\hspace{1cm}
\includegraphics[width=6.5cm,height=6.5cm]{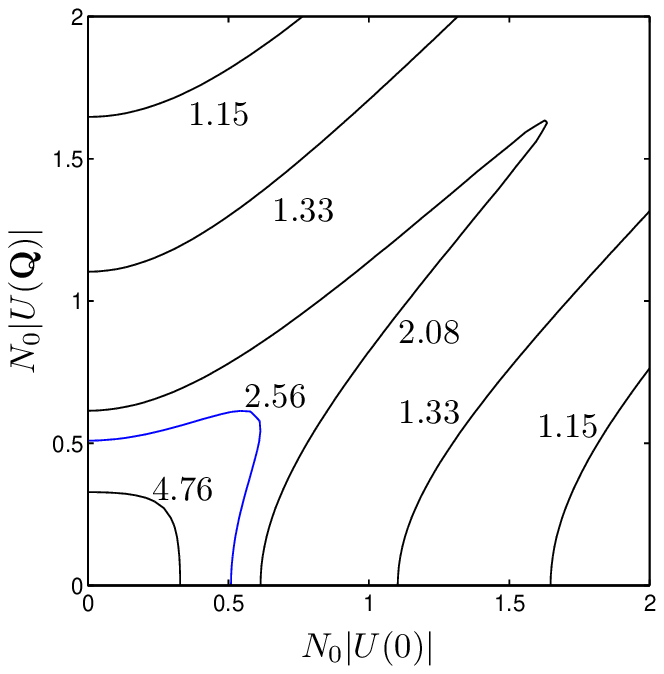}
\end{center}
\caption{\label{kemiai-kontur}(Color online) Left panel: the shift in the chemical potential due to impurity
scattering as a function of the scattering amplitudes. Right panel: contourplot of the scattering rate
$\Gamma/\Gamma_c=1$ on the $U(0)-|U(\mathbf{Q})|$ plane for $n_i/n_c=1.15$, 1.33, 2.08, 2.56 and 4.76. For example
in the case of $n_i=2.56n_c$, the variation of the scattering parameters is restricted to the region inside the
contour (blue) with the corresponding label.}
\end{figure}

Using the off-diagonal component of the Green's function in Eq.~\eqref{exactgreen}, the gap equation, that
determines the temperature dependence of the order parameter in the impure system, reads as
\begin{equation}
\Delta=\frac{PT}{N}\sum_{\mathbf{k},\omega_n}\frac{\Delta\sin(bk_y)^2}{\tilde\omega_n^2+\xi(\mathbf{k})^2+\Delta^2\sin(bk_y)^2},\label{gapegyenlet}
\end{equation}
where $N$ denotes the number of sites. The effective coupling $P$ favoring the sine dependence of
$\Delta(\mathbf{k})$, and a factor of $\sin(bk_y)$ in the numerator comes from the kernel of the electron-electron
interaction.\cite{balazs-sdw} Now we derive the equation that determines the transition temperature
$T_c$. Following Ref.~\onlinecite{roshen}, we reformulate Eq.~\eqref{gapegyenlet} as
\begin{equation}
\ln\left(\frac{T}{T_{c0}}\right)=
T\sum_{\omega_n}\left\{
\frac{4\sqrt{\tilde\omega_n^2}}{\Delta^2}
\left(E\left(\frac{i\Delta}{\tilde\omega_n}\right)-K\left(\frac{i\Delta}{\tilde\omega_n}\right)\right)-\frac{\pi}{|\omega_n|}\right\},\label{uj-gap}
\end{equation}
where $E(z)$ is the complete elliptic integral of the second kind, and $T_{c0}$ is the transition temperature of
the pure UDW.\cite{balazs-sdw} We note that the basic Eqs.~\eqref{selfenergy} and~(\ref{alapegyenlet}-\ref{kisg})
determining the renormalized frequency as a function of the bare frequency provide unphysical solutions as
well. The physically correct solution, that involves the right analytic properties of the Green's function, is
chosen according to: $\text{sign}(\text{Re}\tilde\omega_n)=\text{sign}(\omega_n)$. In the region near the
transition temperature $T_c$, where $\Delta$ is small, we can expand the right hand side of Eq.~\eqref{uj-gap}, and
we obtain the celebrated Abrikosov-Gor'kov (AG) formula
\begin{equation}
-\ln\left(\frac{T_c}{T_{c0}}\right)=\text{Re}\psi\left(\frac12+\frac{|\Sigma_0''|}{2\pi T_c}+i\frac{\Sigma_0'-\delta\mu}{2\pi T_c}\right)+\psi\left(\frac12\right),\label{ag}
\end{equation}
where $\psi(z)$ is the digamma function and
\begin{align}
\Sigma_0'&\equiv\text{Re}\Sigma(\Delta=0)\notag\\
&=4n_iU(0)D^{-1}(4+N_0^2\pi^2(U(0)^2-|U(\mathbf{Q})|^2)),\label{kemiai}\\
\Sigma_0''&\equiv\text{Im}\Sigma(\Delta=0)\notag\\
&=-2\pi\,\text{sign}(\omega_n)n_iN_0D^{-1}[4(U(0)^2+|U(\mathbf{Q})|^2)+\pi^2N_0^2(U(0)^2-|U(\mathbf{Q})|^2)^2],\label{kepzetessigma}\\
D&=(4+N_0^2\pi^2(U(0)+|U(\mathbf{Q})|)^2)(4+N_0^2\pi^2(U(0)-|U(\mathbf{Q})|)^2).\label{nagyd}
\end{align}
Up to this point the shift in chemical potential played the role of a constant parameter, the value of which is
determined by particle conservation via the integrated density of states. However, the calculation and analysis of
the one particle density of states is the subject of the next section, therefore we anticipate here the result
\begin{equation}
\delta\mu=\Sigma_0'.\label{eloleg}
\end{equation}
This result could have been presumed also from the AG formula in Eq.~\eqref{ag}, because in the Born and resonant
limits the argument of $\psi$ is purely real.\cite{balazs-born,balazs-unitary} In Fig.~\ref{kemiai-kontur} (left
panel) we plot $\delta\mu$ as a function of the scattering parameters. Note also, that the above formula is valid
for normal metals with $\Delta=0$ as well. With all this, we arrive to the final form
\begin{equation}
-\ln\left(\frac{T_c}{T_{c0}}\right)=\psi\left(\frac12+\frac{\Gamma}{2\pi
 T_c}\right)+\psi\left(\frac12\right),\label{finalag}
\end{equation}
where we have introduced the scattering rate $\Gamma=|\Sigma_0''|$. We see now, and shall point it out, that the AG
formula is valid in general for arbitrary strength of the impurity scattering, and its applicability is not limited
for the case of weak and strong scatterers only. The general form for $\Gamma$ simplifies to the correct expression
$(\Gamma_1+\Gamma_2)/2$ in the Born limit, where $\Gamma_1=\pi n_iN_0U(0)^2$ and $\Gamma_2=\pi
n_iN_0|U(\mathbf{Q})|^2$ are the forward and backward scattering rates, respectively.\cite{balazs-born} On the
other hand, it gives back the result $2n_i/(\pi N_0)$ known for strong scatterers.\cite{balazs-unitary} Its
critical value, where $T_c$ vanishes, is given by $\Gamma_c=\pi T_{c0}/(2\gamma)=\sqrt{e}\Delta_{00}/4$, with
$\gamma=1.781$ being the Euler constant and $\Delta_{00}$ stands for the pure zero temperature order parameter. One
can easily see that $\Gamma$ is bounded as a function of the forward and backward scattering parameters, but scales
with $n_i$. Consequently, this enables us to define a critical impurity concentration
$n_c=\pi^2N_0T_{c0}/(4\gamma)$, below which $(n_i<n_c)$ the values of $U(0)$ and $|U(\mathbf{Q})|$ can be
arbitrary, thereby including the Born and unitary limits as well, but above which $(n_i>n_c)$ the amplitudes are
constrained into a finite region on the $U(0)-|U(\mathbf{Q})|$ plane around the origin. This is because otherwise
the strength of the impurity scattering completely destroys the density wave condensate even at zero
temperature. In the latter case apparently, only the Born limit is accessible.  In Fig.~\ref{kemiai-kontur}
(right-panel) we show these regions and the corresponding confining contours of $\Gamma/\Gamma_c$. Using the
parameters of $\alpha$-(BEDT-TTF)$_2$KHg(SCN)$_4$, namely $T_c=10$K, $v_F=6\times 10^4$m/s,\cite{balazs-eurlett} and
lattice constant in the chain direction $a=10^{-9}$m,\cite{endo} the critical concentration is estimated as
$n_c=0.01$. Close to $\Gamma_c$ the transition temperature vanishes like\cite{suzumura}
\begin{equation}
\frac{T_c}{T_{c0}}=\frac{\sqrt{6}}{2\gamma}\ln^\frac12\left(\frac{\Gamma_c}{\Gamma}\right).\label{tceltunik}
\end{equation}

\begin{figure}
\includegraphics[width=8.6cm,height=6.5cm]{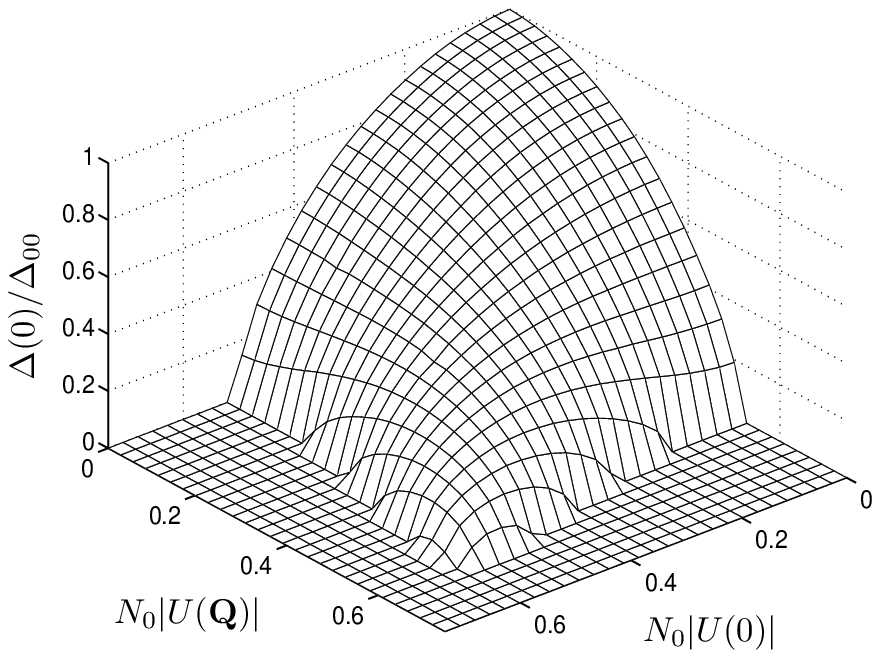}\hfill
\includegraphics[width=8.6cm,height=6.5cm]{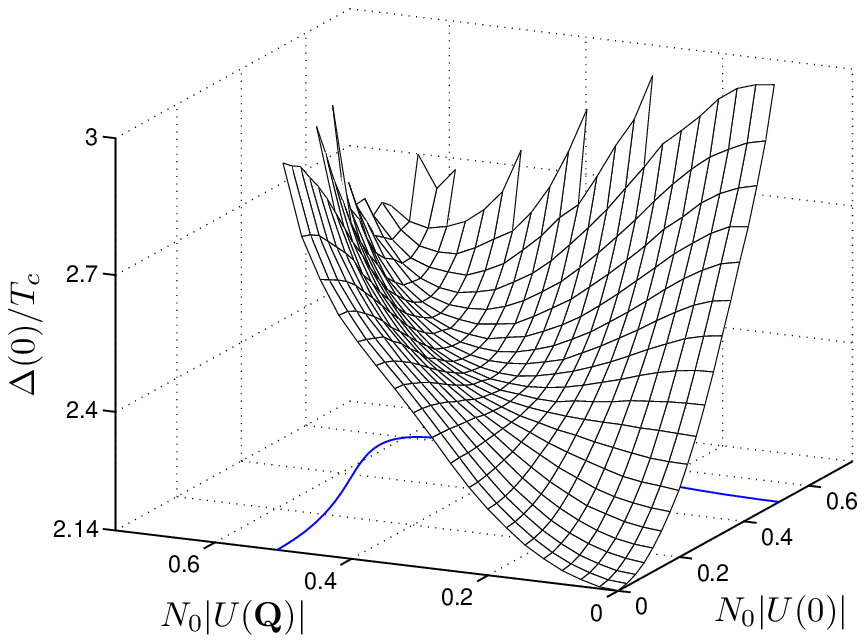}
\caption{\label{order-dptc}(Color online) Left panel: the zero temperature order parameter $\Delta(0)$ as a
function of the scattering amplitudes for fixed impurity concentration $n_i/n_c=2.56$. Right panel: the
$\Delta(0)/T_c$ ratio versus the scattering amplitudes for the same concentration as in the left panel. It is clear
that the presence of impurities enhances this ratio from the pure result 2.14. For both panels, the contour of the
domain on the $U(0)-|U(\mathbf{Q})|$ plane is the blue curve in Fig.~\ref{kemiai-kontur} right panel.}
\end{figure}

Next, we proceed with the analysis of the zero temperature gap maximum. For brevity, in this paragraph
$\Delta\equiv\Delta(T=0)$. Similar calculations that lead to Eq.~\eqref{uj-gap},\cite{roshen,zittartz1} give in our
case
\begin{equation}
\frac{\pi}{2}\ln\left(\frac{\Delta}{\Delta_{00}}\right)=\int_{-\infty}^\infty\frac{d\omega}{\Delta}\left(F\left(\frac{\tilde\omega}{\Delta}\right)-F\left(\frac{\omega}{\Delta}\right)\right),\label{nullaorder}
\end{equation}
where $\omega$ and $\tilde\omega$ are, respectively, the bare and the renormalized Matsubara frequencies (we omitted
the subscript $n$), and
\begin{equation}
F(z)=\sqrt{1+z^2}E\left(\frac{1}{\sqrt{1+z^2}}\right)-\frac{z^2}{\sqrt{1+z^2}}K\left(\frac{1}{\sqrt{1+z^2}}\right).\label{nagyffuggveny}
\end{equation}
The expression on the right hand side of Eq.~\eqref{nullaorder}, though is not adapted well for the numerical
solution, is however convenient for further analytical calculations.  Namely, it enables us to determine the
behavior of the order parameter close to the critical scattering rate $\Gamma_c$. As the transition temperature
and the gap maximum are closely related quantities, we expect similar tendency found for $T_c$ in
Eq.~\eqref{tceltunik}. Indeed, we obtain
\begin{equation}
\Delta=\frac{4\Gamma_c}{\sqrt{3}}\left(1-\frac{2\pi}{9}\frac{\text{Re}\Lambda}{\Gamma_c}\right)^{-1/2}\ln^\frac12\left(\frac{\Gamma_c}{\Gamma}\right),\label{deltasmall}
\end{equation}
where $\Lambda=N_0\partial\Sigma(g_0)/\partial g$, and $g_0=g(\Delta=0,\omega_n>0)$. As we have just pointed out, for
numerical purposes Eq.~\eqref{nullaorder} shall be reformulated, and the following form is obtained
\begin{equation}
\frac{\pi}{2}\ln\left(\frac{\Delta}{\Delta_{00}}\right)=-2\int_0^{C_0'}dz\,F(z)+\frac{2}{\Delta}\int_{C_0'}^\infty
dz\,F(z)\text{Im}\left(\frac{\partial\Sigma}{\partial
  z}\right)+2C_0''\int_0^1d\lambda\,\text{Im}\left\{F(z)\left(1-\frac{i}{\Delta}\frac{\partial\Sigma}{\partial z}\right)\right\},\label{numerikusdelta}
\end{equation}
where $z=\tilde\omega/\Delta$, $C_0=C_0'+iC_0''$ is the decomposition of $C_0$ into real and imaginary parts, and
in the last integral $z=C_0'+i\lambda C_0''$. The $C_0$ parameter is the value of $\tilde\omega_n/\Delta$ at zero
frequency as usual.\cite{balazs-born,balazs-unitary,sun-maki} Note, that due to the true complex nature of the self
energy $\Sigma$ with finite real part, this generalized $C_0$ acquires a finite imaginary part leading to the
appearance of the third term on the right hand side of Eq.~\eqref{numerikusdelta}. We present the result of the
numerical solution of Eq.~\eqref{numerikusdelta} in Fig.~\ref{order-dptc} (left panel) for fixed impurity
concentration $n_i/n_c=2.56$, that exceeds the critical threshold $n_c$. We shall emphasize, that the impurity
concentration and the scattering parameters enter in the expression for $\Delta$ not through $\Gamma$, as is the
case for (U)DWs\cite{roshen,balazs-born,balazs-unitary} and superconductors\cite{puch-maki} in the Born and unitary
limits and also for $T_c$ in the AG formula (see Eq.~\eqref{finalag}), but enter through $\Sigma$ and
$\delta\mu$. In this respect, the plot shows the dependence of the zero temperature order parameter on the
scattering amplitudes above the $U(0)-|U(\mathbf{Q})|$ plane, with $n_i$ fixed. One can easily see, that the
contour where $\Delta$ vanishes and $\Gamma$ takes on the critical value $\Gamma_c$ (see Eq.~\eqref{deltasmall}),
is just the one appearing in the right panel of Fig.~\ref{kemiai-kontur}, with the corresponding label 2.56.

The dependence of $T_c$ on the scattering amplitudes is qualitatively the same as that of $\Delta(0)$ in
Fig.~\ref{order-dptc}, therefore is not presented here. However, the $\Delta(0)/T_c$ ratio might be of interest,
and is shown in the right panel of Fig.~\ref{order-dptc}. It is clear from the figure, that the ratio increases
with increasing strength of impurity scattering. The lower bound is obviously the pure value
$2\pi/(\gamma\sqrt{e})=2.14$.\cite{balazs-sdw} Moreover, the maximum is reached on the confining contour
\begin{equation}
\max\left.\frac{\Delta(0)}{T_c}\right|_{\Gamma_c}=\sqrt{\frac85}\pi\left(1+\frac85\frac{n_c}{n_i}\right)^{-1/2}.\label{dptc}
\end{equation}
Here we have used Eqs.~\eqref{tceltunik} and~\eqref{deltasmall}. Now it is clear that as $n_i$ increases, the
overall maximum of the gap to $T_c$ ratio is $\sqrt{8/5}\pi=3.97$. This result has been obtained previously
for quasi-one dimensional anisotropic superconductors based on a second order Born calculation.\cite{suzumura}

Close to $T_c$, $\Delta(T)$ vanishes in a square-root manner as does usually in mean-field treatments. In the
series expansion of Eq.~\eqref{uj-gap} for small $\Delta$, the term of order $\Delta^2$ provides
\begin{equation}
\Delta^2=8(2\pi T_c)^2\frac{\rho\psi'(\frac12+\rho)-1}{\frac32\psi''(\frac12+\rho)+\frac{1}{12}\frac{\text{Re}\Lambda}{T_c}\psi'''(\frac12+\rho)}\left(1-\frac{T}{T_c}\right),\label{deltatc}  
\end{equation}
where $\psi^{(n)}(z)$ is the $n$th polygamma function and $\rho=\Gamma/(2\pi T_c)$. The term
$(12T_c)^{-1}\text{Re}\Lambda$ in the denominator simplifies to $\rho/3$ in the Born,\cite{balazs-born} and to
$-\rho/3$ in the resonant unitary\cite{puch-maki} limits.

Now we derive an expression for the order parameter close to absolute zero. The alternative formulas for the gap
equation in Eq.~\eqref{gapegyenlet} presented so far are not really convenient for our present purposes, thus with
the usual contour integration technique we rewrite it as
\begin{align}
\frac{\pi}{4}\ln\left(\frac{\Delta}{\Delta_{00}}\right)=&
\phantom{-}\int_{-\infty}^0\frac{d\omega}{\Delta}\left\{\frac{\omega}{\Delta}\text{Re}\left[K\left(\frac{\Delta}{\omega}\right)-E\left(\frac{\Delta}{\omega}\right)\right]-\text{Re}\left[\frac{u}{\Delta}\left(K\left(\frac{\Delta}{u}\right)-E\left(\frac{\Delta}{u}\right)\right)\right]\right\}\notag\\
&-\int_0^\infty\frac{d\omega}{\Delta}f(\omega)\left\{\text{Re}\left[\frac{u}{\Delta}\left(K\left(\frac{\Delta}{u}\right)-E\left(\frac{\Delta}{u}\right)\right)\right]_{\omega}-\text{Re}\left[\frac{u}{\Delta}\left(K\left(\frac{\Delta}{u}\right)-E\left(\frac{\Delta}{u}\right)\right)\right]_{-\omega}\right\},\label{gapegyenlet3}
\end{align}
where $\Delta$ means $\Delta(T)$, and $u$ is the analytically continued value of $i\tilde\omega_n$, namely
$u=i\tilde\omega_n(i\omega_n\to\omega+i0)$. In addition $f(\omega)$ is the Fermi function and the subscripts
$\pm\omega$ mean that $u$ has to be evaluated at the corresponding frequency. This form will also be used for the
evaluation of the low temperature specific heat shortly. We note that in the pure case, Eq.~\eqref{gapegyenlet3}
simplifies considerably, and one can recognize the result found for pure $d$-wave
superconductors.\cite{won-maki-prb} This is not surprising at all, as the similarity between the thermodynamics of
UDW and $d$-SC are well-known.\cite{balazs-sdw,balazs-born,balazs-unitary,sun-maki} From Eq.~\eqref{gapegyenlet3}
we finally obtain
\begin{equation}
\Delta(T)=\Delta(0)-\frac{2\pi}{3}\frac{a}{1-b}\frac{T^2}{\Delta(0)},\label{kisdelta}  
\end{equation}
where the dimensionless parameters $a$ and $b$ are
\begin{align}
a&=-\Delta(0)\,\text{Re}\left\{\left(1+\frac{\partial\Sigma(u_0)}{\partial u}\right)^{-1}\frac{\partial}{\partial u}E'\left(\frac{\Delta(0)}{u_0}\right)\right\},\label{a}\\
b&=\frac{4}{\pi}\int_{-\infty}^0\frac{d\omega}{\Delta(0)}\text{Re}\left\{(\Sigma-\delta\mu)\left(1+\frac{\partial\Sigma}{\partial u}\right)^{-1}\frac{\partial}{\partial u}E'\left(\frac{\Delta(0)}{u}\right)\right\}.\label{b}
\end{align}
In Eqs.~(\ref{a}-\ref{b}) $u_0=i\Delta(0)C_0$, and the prime means differentiation with respect to the
argument. With cumbersome calculations one can show that Eq.~\eqref{kisdelta} gives back the known expression in
the weak scattering Born limit.\cite{balazs-born} The most important message of the above formula is obviously the
fact that the presence of impurities changes the pure $T^3$ power law behavior\cite{balazs-sdw} and transforms it
to a faster $T^2$ decrease.

\begin{figure}
\includegraphics[width=8.6cm,height=6.5cm]{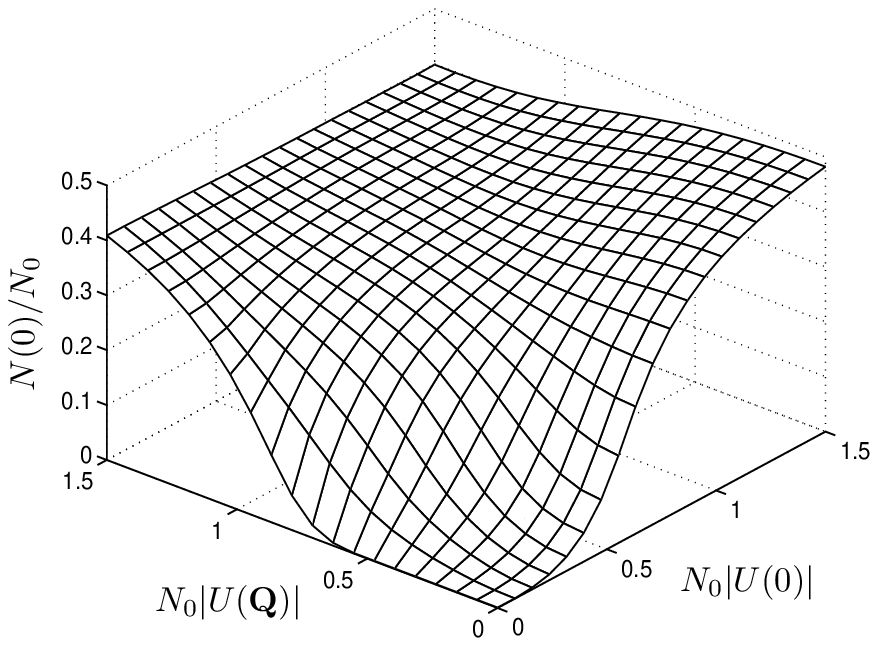}\hfill
\includegraphics[width=8.6cm,height=6.5cm]{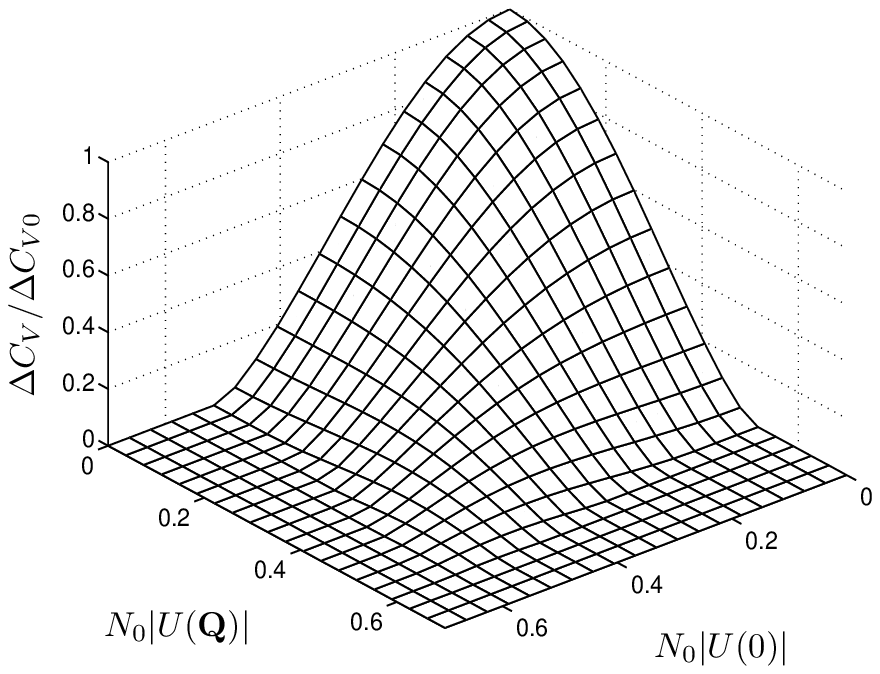}
\caption{\label{fajho}Left panel: the residual density of states at the Fermi level for fixed impurity
concentration $n_i/[N_0\Delta]=0.3$. Right panel: the relative jump in the specific heat at the transition
temperature $T_c$ normalized to the pure value as a function of the scattering amplitudes for fixed impurity
concentration $n_i/n_c=2.56$.}
\end{figure}

We shall close this section with the presentation of some relevant thermodynamic quantities, such as the grand
canonical potential $\Omega$, the entropy $S$ and the specific heat $C_V$. Our starting point is the well-known
formula for $\Omega$ due to Pauli, that is
\begin{equation}
\Omega-\Omega_0=\int_0^1\frac{d\lambda}{\lambda}\langle \lambda H_{int}\rangle,
\label{grandcanonical}
\end{equation}
where $H_{int}$ is the interaction causing the phase transition. This formula gives us the thermodynamic potential
difference $\delta\Omega$ between the normal and the DW phase. After lengthy calculations we obtain
\begin{equation}
\frac{\delta\Omega(T)}{N_0V}=-\frac{\Delta^2}{4}+\frac{\Delta^2}{2}\ln\left(\frac{\Delta}{\Delta_{00}}\right)+
2\int_{-\infty}^0 d\omega\,\omega\left(\frac{N(\omega)}{N_0}-\frac{N_p(\omega)}{N_0}\right)-
4T\int_0^\infty d\omega\ln(1+e^{-\beta\omega})\left(\frac{N^S(\omega)}{N_0}-1\right),\label{potencial}  
\end{equation}
where $\Delta$ means $\Delta(T)$ everywhere. Moreover $N(\omega)$ and $N_p(\omega)$ are, respectively, the
quasiparticle density of states (DOS) in the impure and pure\cite{balazs-sdw} systems, and finally $N^S(\omega)$ is
the symmetrized DOS: $N^S(\omega)=(N(\omega)+N(-\omega))/2$. We will come back to the analysis of DOS in the next
section, and we will see that it is not an even function indeed. We shall emphasize that the order parameter which
is hidden in $N_p(\omega)$ and determines the energy scale, is that of the impure system. We also note that
Eq.~\eqref{potencial} is valid for all temperatures. In obtaining the above formula, we repeatedly made use of
Eq.~\eqref{gapegyenlet3}, and the following important relation
\begin{equation}
\frac{\partial}{\partial\omega}\text{Re}\left\{E'\left(\frac{\Delta}{u}\right)\right\}=\frac{\pi}{2}\frac{\partial}{\partial\Delta}\frac{N(\omega)}{N_0}.  
\end{equation}
Now, one can easily check, that the temperature dependence through $\Delta$ does not contribute to the entropy since
the gap equation ensures $\partial\delta\Omega/\partial\Delta=0$. Consequently, in an UDW we simply get
\begin{equation}
S=4\int_0^\infty d\omega N^S(\omega)\left(\ln(1+e^{-\beta\omega})+\beta\omega f(\omega)\right),\label{entropia}
\end{equation}
which is the standard expression of a normal metal, except the DOS is that of the impure unconventional density
wave. Note that due to the asymmetric feature of the DOS with respect to the Fermi energy, here $N^S(\omega)$ shows
up. In those systems however, where the particle-hole symmetry is not violated, this feature remains hidden
obviously.\cite{maki-puch-pwave} From this, the low temperature specific heat is $C_V=(2\pi^2/3)N(0)T$. Here $N(0)$
stands for the residual density of states on the Fermi level and it is shown in the left panel of Fig.~\ref{fajho}.

Around $T_c$ the jump in the specific heat, normalized to the pure result, follows as
\begin{equation}
\frac{\Delta C_V}{\Delta C_{V0}}=-21\zeta(3)\frac{T_c}{T_{c0}}\frac{\left(1-\rho\psi'(\frac12+\rho)\right)^2}{\frac32\psi''(\frac12+\rho)+\frac{1}{12}\frac{\text{Re}\Lambda}{T_c}\psi'''(\frac12+\rho)},\label{fajhougras}  
\end{equation}
where $\Delta C_{V0}=16\pi^2 N_0T_{c0}/(21\zeta(3))$, and it is shown in the right panel of Fig.~\ref{fajho} for
fixed impurity concentration $n_i=2.56n_c$. This result, just like Eq.~\eqref{deltatc}, also becomes simpler in the
case of weak and strong scattering limits, $(12T_c)^{-1}\text{Re}\Lambda\to\pm\rho/3$, and we reobtain the familiar
results.\cite{puch-maki} Note that in Ref.~\onlinecite{puch-maki} a factor of $T_c/T_{c0}$ is missing, but is
corrected by the same authors in a separate paper.\cite{maki-puch-isopwave}

\section{Density of states in UDW}
With the knowledge of the Green's function (see Eq.~\eqref{exactgreen}), the quasiparticle density of states can be
calculated as
\begin{align}
N(\omega)&\equiv -\frac{1}{\pi V}\sum_\mathbf{k}\text{Im}\,\text{Tr}\left(G^R(\mathbf{k},\omega)\right)\notag\\
&=\frac{2}{\pi}N_0\,\text{Im}\left(\frac{u}{\sqrt{\Delta^2-u^2}}K\left(\frac{\Delta}{\sqrt{\Delta^2-u^2}}\right)\right)=\frac{2}{\pi}N_0\,\text{Re}K\left(\frac{\Delta}{u}\right),\label{dos}  
\end{align}
where $u=i\tilde\omega_n(i\omega_n\to\omega+i0)$ is the analytically continued value of the dressed Matsubara
frequency. It is determined self-consistently from
\begin{equation}
u=\omega+\delta\mu-\Sigma(u),\label{kifolytatott}
\end{equation}
and as we have already pointed out before, the physically correct solution is the one that ensures the right analytic
properties of the retarded Green's function, that is $\text{Im}u>0$.

Now, we turn our attention to the shift in the chemical potential appearing first in Eq.~\eqref{alapegyenlet}, and
here in the analytically continued version Eq.~\eqref{kifolytatott} as well. We have noted in the previous section
also, that $\delta\mu$ was introduced by hand in order to maintain particle conservation. Its actual value is to be
determined from the integrated DOS as
\begin{equation}
\int_{-\infty}^0 d\omega(N(\omega)-N_0)=0.\label{muegyenlet}  
\end{equation}
The second term in the integrand is the constant density of states of the pure metal, but it could equally be the
DOS of the pure UDW. This is because the metal to UDW transition without impurities conserves the particle-hole
symmetry and the even parity of the DOS, and as such, conserves particle number. The numerical solution of
Eq.~\eqref{muegyenlet} confirms our expectation (based on the AG formula, see Eq.~\eqref{ag}), and we indeed find
$\delta\mu=\Sigma_0'$.

\begin{figure}
\includegraphics[width=8.6cm,height=6.5cm]{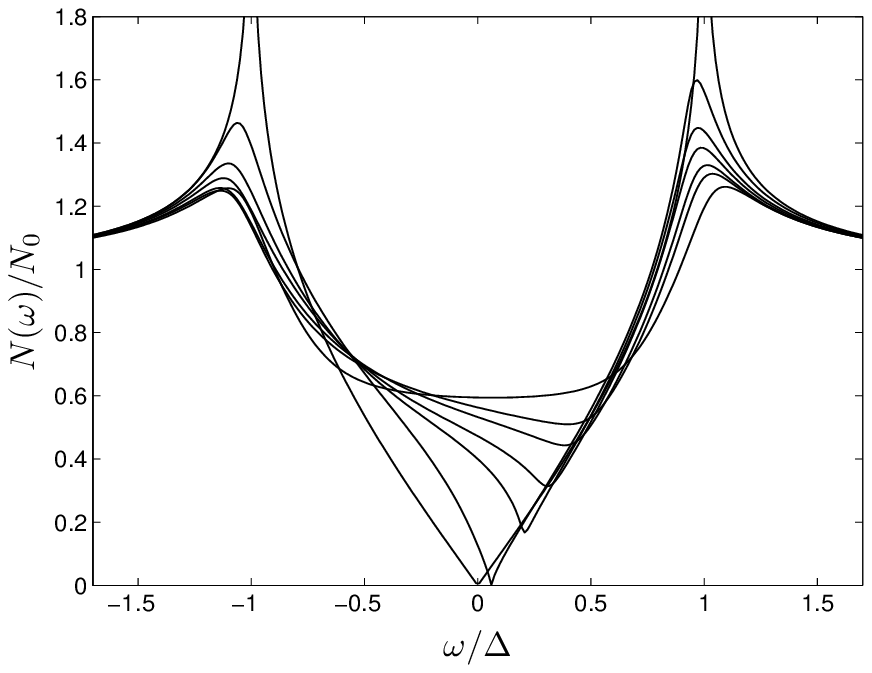}\hfill
\includegraphics[width=8.6cm,height=6.5cm]{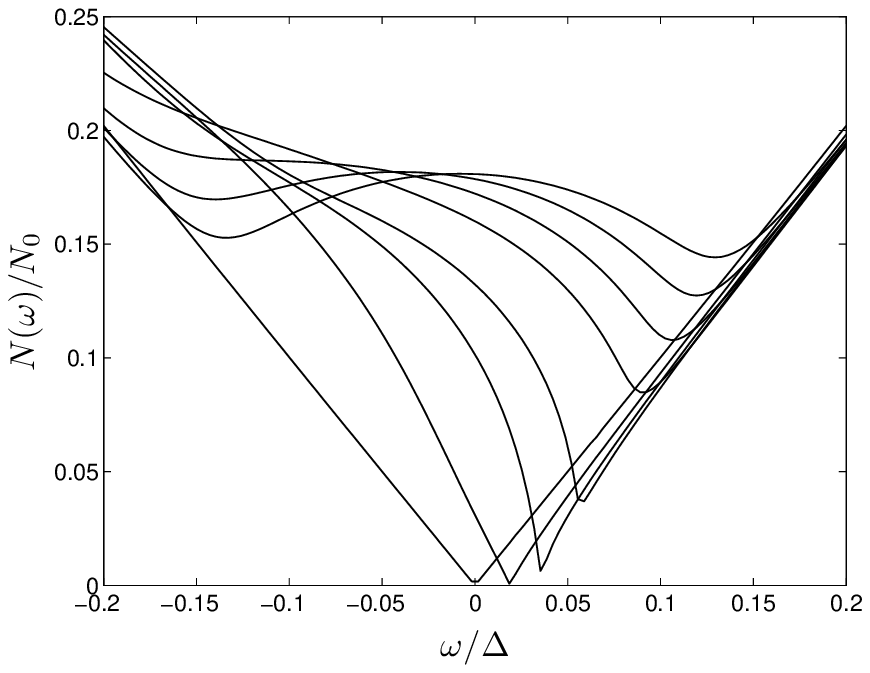}
\caption{\label{dosabra}Left panel: the quasiparticle density of states as a function of reduced energy for
different values of the scattering amplitudes at fixed impurity concentration
$n_i/[N_0\Delta]=0.4$. $U(0)/|U(\mathbf{Q})|=2$ and $N_0|U(\mathbf{Q})|=0$, 0.2, 0.4, 0.6, 1, 1.5, and 20 for
curves with increasing $N(0)$. Right panel: the density of states enlarged around the Fermi energy for
$n_i/[N_0\Delta]=0.02$, the ratio of the two scattering parameters is the same as in the left panel and
$N_0|U(\mathbf{Q})|=0$, 0.6, 1, 1.5, 3, 5, 10, and 50 for curves with increasing $N(0)$.}.
\end{figure}

With all this, we are now able to numerically evaluate the quasiparticle DOS defined in Eq.~\eqref{dos}, and the
results are shown in Fig.~\ref{dosabra} for positive, i.e. repulsive forward scattering. Apparently, the main
feature of these curves is the violation of particle-hole symmetry, similar to that found in $d$-density
waves\cite{d-density-wave-bdg}. Namely, strictly speaking they are not even functions of energy. However, in the
weak and strong scattering limits we reobtain even parity.\cite{suzumura,balazs-born,balazs-unitary} In addition,
it is clear from the figures, that the logarithmic singularities at $\omega=\pm\Delta$ characteristic to UDWs in
general are smoothed out due to the presence of scatterers. In the case of repulsive (attractive) scatterers the
upper (lower) peak is larger. An other well-known effect of impurities is the finite residual DOS at the Fermi
energy, $N(0)>0$. For small forward and backward scatterings it disappears exponentially, it is shown separately in
the left panel of Fig.~\ref{fajho}. The right panel of Fig.~\ref{dosabra} demonstrates the behavior of DOS on a
smaller energy scale around the Fermi energy. For small $n_i$ and large scattering amplitudes, we come up with the
familiar result of the resonant limit,\cite{balazs-unitary,sun-maki,hotta} that the density of states exhibits an
island of localized states around the Fermi energy superimposed on the usual gapless density of states of the pure
system, which is manifested in the nonmonotonic nature of the DOS close to $\omega=0$.

After some algebra, the low-energy behavior turns out to be linear, and it reads
\begin{equation}
N(\omega)=N(0)+\frac{2}{\pi}N_0\,\text{Im}\left(\frac{K-E}{\sqrt{1+C_0^2}}\left(1+\frac{\partial\Sigma(u_0)}{\partial
u}\right)^{-1}\right)\frac{\omega}{\Delta},
\end{equation}
where the argument of the elliptic integrals is $1/\sqrt{1+C_0^2}$. This is to be contrasted with the $\omega^2$
contribution of the extreme limits. On the other hand, at high energies the density of states reaches the normal
state value as
\begin{equation}
N(\omega)=N_0\left(1+\frac{\Delta^2}{4}\frac{\omega^2-\Gamma^2}{(\omega^2+\Gamma^2)^2}\right).
\end{equation}
This is what we find in the second order Born calculation,\cite{balazs-born} except that now $\Gamma$ is defined by
the general formula $\Gamma=|\Sigma_0''|$ (see also Eqs.~(\ref{kepzetessigma}-\ref{nagyd})). It is valid for
arbitrary strength of impurity scattering parameters.

\section{Conclusion}
We have studied the effect of non-magnetic impurities in quasi-one dimensional unconventional charge and spin
density waves within mean-field theory. The focus has been on the thermodynamic response of the system. As the
impurities are of the ordinary non-magnetic type, the spin degrees of freedom in the condensate play no role,
leading to the same conclusions for both types of UDW. As opposed to the usual treatment of impurity scattering in
density waves and superconductors (i.e. the Born and unitary limits), we do not put any constraint on the strength
of forward and backward scattering amplitudes, but make use of the full self energy obtained in the self-consistent
non-crossing approximation (NCA). Namely, the applied proper self energy includes all ladder-, and rainbow-type
diagrams to infinite order. In this regard, our NCA is full. We than repeatedly took advantage of the fact, that
our generalized formulas contain the familiar results of the weak and strong scattering limits, and we reobtained
them indeed. On the other hand, our results provide explicit expressions in the intermediate regime as well.

We started our investigation with the transition temperature and the order parameter affected by the scatterers. We
have found that, with a suitable redefinition of the scattering rate compared to either the Born or the unitary
expressions, the Abrikosov-Gor'kov formula holds in this general case as well. Its validity can be therefore
extended to arbitrary scattering amplitudes. The analysis of the AG formula allows one to introduce a critical
impurity concentration, below which $(n_i<n_c)$ the impurity potential, whatever strong it is, cannot break down
the density wave condensate. However, if $n_i$ exceeds the critical threshold $n_c$, the order of which is
estimated around $1\%$, the scattering amplitudes cannot be arbitrary large and only the Born limit is accessible.
An interesting consequence of impurity presence is the appearance of a shift in the chemical potential $\delta\mu$,
the inclusion of which is necessary in order to conserve particle number. We gave an explicit and exact expression
for $\delta\mu$ as a function of the scattering amplitudes and concentration, that is valid in a normal metal as
well. The overall maximum of the zero temperature gap to $T_c$ ratio was found to be the same
$(\sqrt{8/5}\pi=3.97)$ as in anisotropic quasi-one dimensional superconductors.

The most eye-catching feature of impurities is definitely the violation of particle-hole symmetry manifested in the
asymmetric quasiparticle density of states. Strictly speaking, the DOS is not an even function of energy with
respect to the proper Fermi energy, that is the pure Fermi energy shifted by $\delta\mu$. The asymmetry is most
apparent around $\omega=0$, where the low frequency behavior is linear as opposed to the quadratic form of the
extreme limits. This also involves that the minimum of DOS is not at the Fermi energy. In addition, the peaks
around $\omega=\pm\Delta$, the residues of logarithmic singularities of the pure DW, become asymmetric as well: in
case of repulsive (attractive) scatterers the upper (lower) peak is larger. In unconventional DWs, at any finite
scattering strength the valley of the DOS around the Fermi energy is filled in, leading to normal electronlike
behavior very close to absolute zero, but the reduced density of states compared to the normal state signals the
effect of the condensate. Taking the weak and strong scattering limits of the DOS, one regains particle-hole
symmetry, and can easily reobtain the familiar lineshapes of the Born and unitary limits, in particular for
instance the island of localized states around $\omega=0$ in the latter case.

\begin{acknowledgments}
This work was supported by the Hungarian National Research Fund under grant numbers OTKA NDF45172, T046269,
TS049881. B. D. was supported by the Magyary Zolt\'an postdoctoral program of Magyary Zolt\'an Foundation for
Higher Education (MZFK).
\end{acknowledgments}

\bibliography{impurity2}

\end{document}